# Determination of anisotropic dipole moments in self-assembled quantum dots using Rabi oscillations

A. Muller, Q. Q. Wang,[a)] P. Bianucci, and C. K. Shih[b)]

Department of Physics, University of Texas at Austin, Austin, Texas 78712

Q. K. Xue

International Center for Quantum Structures (ICQS), Institute of Physics, The Chinese Academy of Sciences, Beijing 100080, P. R. China

By investigating the polarization-dependent Rabi oscillations using photoluminescence spectroscopy, we determined the respective transition dipole moments of the two excited excitonic states $|E_x\rangle$ and $|E_y\rangle$ of a single self-assembled quantum dot that are nondegenerate due to shape anisotropy. We find that the ratio of the two dipole moments is close to the physical elongation ratio of the quantum dot.

Recently, semiconductor quantum dots (QDs) have emerged as potential key elements in the emerging field of quantum information processing. Not only do QDs possess optical properties similar to atoms,[1] but they also benefit from well-established semiconductor manufacturing capabilities making large-scale device integration possible. Indeed, much effort is now devoted to a possible implementation of elementary quantum computation in QD systems,[2] which was especially stimulated by the observation of Rabi oscillations (ROs) of excitonic states by several groups[3,4,5] within the past two years. These oscillations that the population of an excited state undergoes as a function of the input pulse area, or qubit rotations, are the hallmark of a coherent system interacting with the electromagnetic field. But, besides marking a significant milestone, recording of ROs is also a direct way of measuring dipole moments of excitonic transitions which are much larger than their atomic counterparts. These can provide more insight about the fundamental structure of QDs which yet remains largely unexplored. Previous studies[4] have found dipole moments typically around 20 Debye in self-assembled QDs where the confinement is strong and dimensions are small compared to interface fluctuation QDs (IFQDs) where dipole moments close to 100 Debye have been measured.[5] An individual excitonic electric dipole transition approximates a two-level system so that many coherent phenomena such as ROs can be described by the familiar Bloch equations of atomic physics.[6] However, there are interesting properties unique to QDs, reflecting their macroscopic size of thousands of atoms. In particular, shape (and composition) anisotropy leads to a fine structure splitting of the excitonic state, which has been examined directly through photoluminescence excitation (PLE) studies[7] and examined more indirectly using two-pulse excitation where a beating pattern appears in the wavefunction autocorrelation function evidencing that two states rather than one are excited.[8] The long range part of the electron hole exchange interaction is responsible for the splitting[9] into two states that can be selectively excited by appropriate linear polarization of the exciting field.

Here we examine the consequence of shape anisotropy in the ROs of an excited excitonic state using photoluminescence (PL) spectroscopy. We focused on a single asymmetric QD that shows a large fine structure splitting and recorded Rabi oscillations by bringing the laser into resonance with either of the two transitions. We find that the dipole moments measured directly reflect the physical structure of the QD. More precisely, for all QDs examined, a larger dipole moment was measured when the laser polarization was along the elongated axis of the QD.

Our sample contains $In_{0.5}Ga_{0.5}As$ QDs grown by molecular beam epitaxy.[10] The QDs have an average lateral size, height and dot-to-dot separation of 20-40, 4.5 and 100 nm, respectively. Cross-sectional scanning tunneling microscope studies (XSTM) revealed a truncated pyramidal shape with the base in the (0 0 1) plane.[11] The QDs were also found to be elongated in the [1 1 0] direction (which we shall call the *y*-axis throughout) with an average elongation ratio of 1.36 with respect to the [-1 1 0] direction (*x*-axis throughout). The sample was excited with a Ti:Sapphire laser providing 6 ps long pulses at a repetition rate of 80 MHz. To achieve single QD spectroscopy, the laser energy was tuned so as to excite only the small fraction of QDs that have an optically allowed transition at that particular energy. Because decay of excited excitonic states is largely dominated by phonon assisted nonradiative relaxation to the QD ground states, measurement of the PL intensity from recombination of (ground state) excitons monitors the population of the excited state excitons. This PL is emitted at a different energy for each QD (highly inhomogeneous size distribution among QDs) so that each QD can be accessed individually simply by using an imaging spectrograph and a charge-coupled device (CCD) array detector.[12]

In this manner, the PLE spectrum for one particular QD was recorded and is shown in Fig. 1(a) for $x$ (squares) and $y$-polarized excitation (triangles). The two Lorentzian resonances are attributed to the $\Pi_x$ and $\Pi_y$ transitions from the ground state $|0\rangle$ to the split excited states $|E_x\rangle$ and $|E_y\rangle$, respectively, as depicted in the energy diagram of Fig. 1(b). Note that the width of the resonances is limited by the spectral width of the exciting laser (~250 μeV). Their splitting of $\Delta=E_x-E_y \sim 70$ μeV is somewhat larger than those measured in IFQs.[7] Note that the $|E_x\rangle$ state (along the short axis of the QD) is at a larger energy than the $|E_y\rangle$ state as one would expect, because a smaller dimension implies larger confinement energy. Nevertheless, previous studies found that the polarity of the splitting need not be directly correlated to the shape of the QD. Rather, it depends on the number of nodes of the exciton's wavefunction and is generally positive (i.e. $E_x>E_y$) for a first excited state,[7,9] in agreement with the relaxation energy (difference between ground state and excited state energy) of only ~15 meV for that particular QD.

Next, we recorded ROs of states $|E_x\rangle$ and $|E_y\rangle$ with the laser incident on the $x/y$ plane of the sample. The angle of incidence was ~68°. The PL was collected normal to that surface with conventional far-field optics. In Fig. 2, the PL was recorded while the average laser intensity was varied for $x$ (squares) and $y$ (triangles) linear polarization of the laser. Note that for each series, the laser was tuned to the resonance frequency of each transition so that the detuning was zero in both cases. The PL, which is proportional to the population of the excited state, undergoes sinusoidal oscillations as a function of the time-integrated Rabi frequency, or input pulse area $\vartheta(t) = (\mu/\hbar)\int_{-\infty}^{t} E(\tau)d\tau$, where $\mu$ and *E(t)* are the transition dipole moment and electric field envelope, respectively. $\vartheta(\infty)$ is proportional to the square root of the average intensity of the laser. The first maximum corresponds to a π-pulse ($\vartheta=\pi$) that, if dephasing is neglected, leaves the two-level system entirely in the excited state. The first minimum corresponds to a 2π pulse, during which the system has been cycled through the excited state and entirely back to the ground state. At any intermediate intensity, the exciting pulse leaves the system in a superposition state. The results are described well by integrating the optical Bloch equations for a two-level system[6] and plotting the population of the upper state at the end of the pulse versus the input pulse area (solid curves in Fig. 2). Also note that the oscillations are damped due to input pulse area-dependent excitonic damping terms.[13] From the periodicity of the oscillations, transition dipole moments of $\mu_x$=17 Debye and $\mu_y$=22 Debye were extracted for the $\Pi_x$ and $\Pi_y$ transition, respectively. Because of the glancing incidence of the laser, one needs to account for the difference in transmission at the air/GaAs interface for different polarizations. An intensity correction factor of 2.89 was obtained from reflectivity measurements and is in agreement with calculations using the Fresnel equations for reflection and refraction[14] using a refractive index *n*=3.6 for GaAs. The larger dipole moment corresponds to the elongated axis of the QD as expected and the ratio of the two dipole moments is $\mu_y/\mu_x$=1.29. Three other QDs yield ratios $\mu_y/\mu_x$ of 1.24, 1.35 and 1.60 corresponding to an average for all four QDs of $<\mu_y/\mu_x>$=1.37, close to the average elongation ratio obtained from XSTM studies. This suggests that the transition dipole moment should roughly scale with the physical dimensions of the QD, a result that has been obtained from theoretical calculations.[15] For that reason, the particularly large IFQs have attracted much attention, albeit their weak confining potential, for applications where large Rabi frequencies, i.e. large transition dipole moments are desirable, for example in reaching the strong coupling regime of a QD and a microcavity mode.

In summary, we have shown that QD shape anisotropy can be examined by direct measurement of the transition dipole moments of the nondegenerate doublet of excited states $|E_x\rangle$ and $|E_y\rangle$ that can be selectively excited using *x* and *y* linearly polarized light, respectively. This demonstrates that it is possible to gain information about the physical structure of a single QD by taking advantage of a highly coherent process, namely excitonic Rabi oscillations.

This work was supported by NSF-NIRT (DMR-0210383), NSF-FRG (DMR-0071893, DMR-0306239), Texas Advanced Technology program, and the W.M. Keck Foundation.

**References and Notes**

a) Present Address: International Center for Quantum Structures (ICQS), Institute of Physics, The Chinese Academy of Sciences, Beijing 100080, P. R. China

b) Electronic mail: shih@physics.utexas.edu

**Figure Captions**

FIG. 1. Energy structure of a single, highly asymmetric QD. (a) PLE spectrum (first-excited state) of the $\Pi_x$ (squares) and $\Pi_y$ (triangles) transitions. (b) QD energy diagram. The split states are denoted by $|E_x\rangle$ and $|E_y\rangle$. The QD ground state $|1'\rangle$ is a spectator state used to monitor the population of $|E_x\rangle$ and $|E_y\rangle$.

FIG. 2. Rabi oscillations of states $|E_x\rangle$ and $|E_y\rangle$. The PL from the $|1'\rangle \rightarrow |0\rangle$ transition was recorded while the average laser intensity was varied for $x$ (squares) and $y$ (triangles) linearly polarized excitation. The solid line is a fit to the data.

**Figures**

FIG. 1.

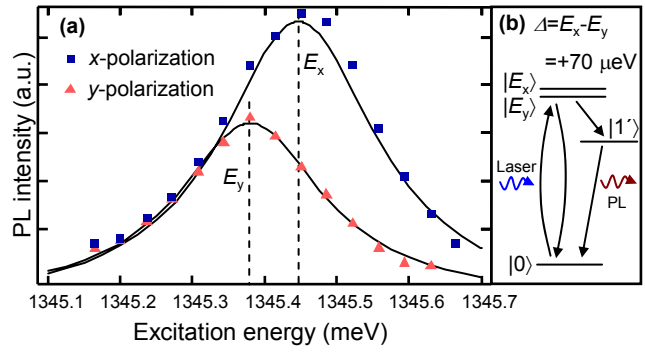

FIG. 2.

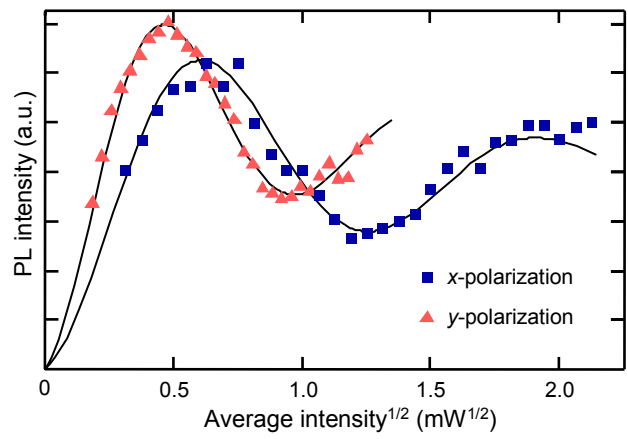